\input harvmac
\input epsf

\newcount\figno
\figno=0
\def\fig#1#2#3{
\par\begingroup\parindent=0pt\leftskip=1cm\rightskip=1cm\parindent=0pt
\baselineskip=12pt
\global\advance\figno by 1
\midinsert
\epsfxsize=#3
\centerline{\epsfbox{#2}}
\vskip 14pt

{\bf Fig. \the\figno:} #1\par
\endinsert\endgroup\par
}
\def\figlabel#1{\xdef#1{\the\figno}}
\def\encadremath#1{\vbox{\hrule\hbox{\vrule\kern8pt\vbox{\kern8pt
\hbox{$\displaystyle #1$}\kern8pt}
\kern8pt\vrule}\hrule}}

\overfullrule=0pt

\noblackbox
\parskip=1.5mm

\def\Title#1#2{\rightline{#1}\ifx\answ\bigans\nopagenumbers\pageno0
\else\pageno1\vskip.5in\fi \centerline{\titlefont #2}\vskip .3in}

\font\caps=cmcsc10

\noblackbox
\parskip=1.5mm



   \def\CG{{\cal G}}
   
  \def\CD{{\cal D}} 
   
\def\CN{{\cal N}}


\def\dj{\hbox{d\kern-0.347em \vrule width 0.3em height 1.252ex depth
-1.21ex \kern 0.051em}}

\def\Tr{{\rm Tr\,}}

\def\ket{\rangle}
\def\bra{\langle}

\def\pt{\partial}

\def\Dirac{\,\raise.15ex\hbox{/}\mkern-13.5mu D}
\def\dirac{\,\raise.15ex\hbox{/}\kern-.57em \partial}
\def\aslash{\,\raise.15ex\hbox{/}\mkern-13.5mu A}

\def\shalf{{\ifinner {\textstyle {1 \over 2}}\else {1 \over 2} \fi}}
\def\sshalf{{\ifinner {\scriptstyle {1 \over 2}}\else {1 \over 2} \fi}}
\def\sfourth{{\ifinner {\textstyle {1 \over 4}}\else {1 \over 4} \fi}}
\def\sthreehalfs{{\ifinner {\textstyle {3 \over 2}}\else {3 \over 2} \fi}}
\def\sdhalfs{{\ifinner {\textstyle {d \over 2}}\else {d \over 2} \fi}}
\def\sdmtwohalfs{{\ifinner {\textstyle {d-2 \over 2}}\else {d-2 \over 2} \fi}}
\def\sdmasonehalfs{{\ifinner {\textstyle {d+1 \over 2}}\else {d+1 \over 2} \fi}}
\def\sdmasthreehalfs{{\ifinner {\textstyle {d+3 \over 2}}\else {d+3 \over 2} \fi}}
\def\sdmastwohalfs{{\ifinner {\textstyle {d+2 \over 2}}\else {d+2 \over 2} \fi}}

 \lref\amps{
   A.~Almheiri, D.~Marolf, J.~Polchinski and J.~Sully,
  ``Black Holes: Complementarity or Firewalls?,''
  JHEP {\bf 1302}, 062 (2013)
  [arXiv:1207.3123 [hep-th]].
  }

\lref\malda{
J.~M.~Maldacena,
  ``Eternal black holes in anti-de Sitter,''
JHEP {\bf 0304}, 021 (2003).
[hep-th/0106112].
}

\lref\us{
J.~L.~F.~Barbon and E.~Rabinovici,
  ``Very long time scales and black hole thermal equilibrium,''
JHEP {\bf 0311}, 047 (2003).
[hep-th/0308063].
}

\lref\BarbonJR{
  J.~L.~F.~Barbon and E.~Rabinovici,
[hep-th/0503144].
}
\lref\BarbonCE{
  J.~L.~F.~Barbon and E.~Rabinovici,
Fortsch.\ Phys.\  {\bf 52}, 642 (2004).
[hep-th/0403268].
}

\lref\ushag{
  J.~L.~F.~Barbon and E.~Rabinovici,
  ``Touring the Hagedorn ridge,''
In *Shifman, M. (ed.) et al.: From fields to strings, vol. 3* 1973-2008.
[hep-th/0407236].
}

\lref\scr{
 Y.~Sekino and L.~Susskind,
  ``Fast Scramblers,''
JHEP {\bf 0810}, 065 (2008).
[arXiv:0808.2096 [hep-th]].
}

\lref\susmal{
 J.~Maldacena and L.~Susskind,
  ``Cool horizons for entangled black holes,''
Fortsch.\ Phys.\  {\bf 61}, 781 (2013).
[arXiv:1306.0533 [hep-th]].
}

\lref\mvr{
  M.~Van Raamsdonk,
  ``Comments on quantum gravity and entanglement,''
[arXiv:0907.2939 [hep-th]].

M.~Van Raamsdonk,
  ``Building up spacetime with quantum entanglement,''
Gen.\ Rel.\ Grav.\  {\bf 42}, 2323 (2010), [Int.\ J.\ Mod.\ Phys.\ D {\bf 19}, 2429 (2010)].
[arXiv:1005.3035 [hep-th]].

M.~Van Raamsdonk,
  ``A patchwork description of dual spacetimes in AdS/CFT,''
Class.\ Quant.\ Grav.\  {\bf 28}, 065002 (2011).
}

\lref\polmarol{
D.~Marolf and J.~Polchinski,
  ``Gauge/Gravity Duality and the Black Hole Interior,''
Phys.\ Rev.\ Lett.\  {\bf 111}, 171301 (2013).
[arXiv:1307.4706 [hep-th]].
}

\lref\deutsch{
J. ~M.~Deutsch,
``Quantum statistical mechanics in a closed system," 
Phys. Rev. A {\bf 43}, 2046 (1991)
}

\lref\peres{
A.~Peres,
``Ergodicity and mixing in quantum theory I,"
Phys. Rev. A {\bf 30}, 504 (1984).

M.~Feingold and A.~Peres,
``Distribution of matrix elements in chaotic systems,"
Phys. Rev. A {\bf 34}, 591 (1986).}

\lref\srednicki{
M.~Srednicki,
``Chaos and quantum thermalization,"
Phys. Rev. E {\bf 50}, 888 (1994) [arXiv:cond-mat/9403051].
 
M.~Srednicki,
``The approach to thermal equilibrium in quantized chaotic systems,"
  J. Phys. A {\bf 32}, 1163 (1999) [arXiv:cond-mat/9809360].
}

\lref\more{
 D.~Birmingham, I.~Sachs and S.~N.~Solodukhin,
  ``Relaxation in conformal field theory, Hawking-Page transition, and quasinormal modes,''
Phys.\ Rev.\ D {\bf 67}, 104026 (2003).
[hep-th/0212308].
M.~Kleban, M.~Porrati and R.~Rabadan,
  ``Poincare recurrences and topological diversity,''
JHEP {\bf 0410}, 030 (2004).
[hep-th/0407192].
}

\lref\festucciaI{
 G.~Festuccia and H.~Liu,
  ``Excursions beyond the horizon: Black hole singularities in Yang-Mills theories. I.,''
JHEP {\bf 0604}, 044 (2006).
[hep-th/0506202].
}

\lref\festucciaII{
G.~Festuccia and H.~Liu,
  ``The Arrow of time, black holes, and quantum mixing of large N Yang-Mills theories,''
JHEP {\bf 0712}, 027 (2007).
[hep-th/0611098].
}

\lref\javi{
J.~L.~F.~Barbon and J.~M.~Magan,
  ``Chaotic Fast Scrambling At Black Holes,''
Phys.\ Rev.\ D {\bf 84}, 106012 (2011).
[arXiv:1105.2581 [hep-th]].
 J.~L.~F.~Barbon and J.~M.~Magan,
  ``Fast Scramblers Of Small Size,''
JHEP {\bf 1110}, 035 (2011).
[arXiv:1106.4786 [hep-th]].
J.~L.~F.~Barbon and J.~M.~Magan,
  ``Fast Scramblers, Horizons and Expander Graphs,''
JHEP {\bf 1208}, 016 (2012).
[arXiv:1204.6435 [hep-th]].
J.~ L.~F.~Barbon and J.~M.~Magan,
  ``Fast Scramblers And Ultrametric Black Hole Horizons,''
JHEP {\bf 1311}, 163 (2013).
[arXiv:1306.3873 [hep-th]].
}

\lref\statedepv{
 E.~Verlinde and H.~Verlinde,
  ``Black Hole Entanglement and Quantum Error Correction,''
JHEP {\bf 1310}, 107 (2013).
[arXiv:1211.6913 [hep-th]].

E.~Verlinde and H.~Verlinde,
  ``Behind the Horizon in AdS/CFT,''
[arXiv:1311.1137 [hep-th]].
}

\lref\statedepp{
 K.~Papadodimas and S.~Raju,
  ``An Infalling Observer in AdS/CFT,''
JHEP {\bf 1310}, 212 (2013).
[arXiv:1211.6767 [hep-th]].

 K.~Papadodimas and S.~Raju,
  ``The Black Hole Interior in AdS/CFT and the Information Paradox,''
Phys.\ Rev.\ Lett.\  {\bf 112}, 051301 (2014).
[arXiv:1310.6334 [hep-th]].
K.~Papadodimas and S.~Raju,
  ``State-Dependent Bulk-Boundary Maps and Black Hole Complementarity,''
[arXiv:1310.6335 [hep-th]].
}

\lref\ellos{
V.~Balasubramanian, M.~Berkooz, S.~F.~Ross and J.~Simon,
  ``Black Holes, Entanglement and Random Matrices,''
[arXiv:1404.6198 [hep-th]].
}

\lref\typi{
 V.~Balasubramanian, B.~Czech, V.~E.~Hubeny, K.~Larjo, M.~Rangamani and J.~Simon,
  ``Typicality versus thermality: An Analytic distinction,''
Gen.\ Rel.\ Grav.\  {\bf 40}, 1863 (2008).
[hep-th/0701122].
}

\lref\beforeamps{
 S.~L.~Braunstein, S.~Pirandola and K.~Zyczkowski,
  ``Better Late than Never: Information Retrieval from Black Holes,''
Phys.\ Rev.\ Lett.\  {\bf 110}, no. 10, 101301 (2013).
[arXiv:0907.1190 [quant-ph]].
}

\lref\mathur{
S.~D.~Mathur,
  ``The Information paradox: A Pedagogical introduction,''
Class.\ Quant.\ Grav.\  {\bf 26}, 224001 (2009).
[arXiv:0909.1038 [hep-th]].
}
\lref\susnew{
 L.~Susskind,
  ``Butterflies on the Stretched Horizon,''
[arXiv:1311.7379 [hep-th]].

L.~Susskind,
  ``Computational Complexity and Black Hole Horizons,''
[arXiv:1402.5674 [hep-th]].

L.~Susskind,
  ``Addendum to Computational Complexity and Black Hole Horizons,''
[arXiv:1403.5695 [hep-th]].
}

\lref\oldsus{
L.~Dyson, M.~Kleban and L.~Susskind,
  ``Disturbing implications of a cosmological constant,''
JHEP {\bf 0210}, 011 (2002).
[hep-th/0208013].
}

\lref\shenstan{
 S.~H.~Shenker and D.~Stanford,
  ``Black holes and the butterfly effect,''
JHEP {\bf 1403}, 067 (2014).
[arXiv:1306.0622 [hep-th]].
S.~H.~Shenker and D.~Stanford,
  ``Multiple Shocks,''
[arXiv:1312.3296 [hep-th]].
}


\baselineskip=15pt

\line{\hfill IFT-UAM/CSIC-14-034}

\vskip 0.5cm

\Title{\vbox{\baselineskip 12pt\hbox{}
 }}
{\vbox {\centerline{Geometry And Quantum Noise
 }
}}

\vskip 0.4cm

\centerline{$\quad$ {\caps
Jos\'e L.F. Barb\'on$^\dagger$
 and
Eliezer Rabinovici$^{\star, c, p}$
}}
\vskip0.5cm

\centerline{{\sl  $^\dagger$ Instituto de F\'{\i}sica Te\'orica IFT UAM/CSIC }}
\centerline{{\sl  C/ Nicol\'as Cabrera 13,
 Campus Universidad Aut\'onoma de Madrid}}
\centerline{{\sl  Madrid 28049, Spain }}
\centerline{{\tt jose.barbon@csic.es}}

\vskip0.1cm

\centerline{{\sl $^\star$
Racah Institute of Physics, The Hebrew University }}
\centerline{{\sl Jerusalem 91904, Israel}}
\centerline{{\tt eliezer@vms.huji.ac.il}}

\vskip0.1cm

\centerline{{\sl $^c$
Kavli Institute for Theoretical Physics, 
University of California}}
\centerline{{\sl Santa Barbara, CA 93106-4030 }}

 \vskip0.1cm
 
\centerline{{\sl $^p$
Laboratoire de Physique Th\'eorique et Hautes Energies,
Universit\'e Pierre et Marie Curie}}
\centerline{{\sl 4 Place Jussieu, 75252 Paris Cedex 05, France}}

\vskip0.5cm

\centerline{\bf ABSTRACT}

 \vskip 0.3cm

 \noindent

We study the fine structure of long-time quantum noise in correlation functions of AdS/CFT systems. Under standard assumptions of quantum chaos for the dynamics and the observables,  we estimate the size of exponentially small oscillations and trace them back to geometrical
features of the bulk system. The noise level is highly suppressed by  the amount of dynamical chaos and the amount of  quantum impurity in the states. This implies that, despite their missing on the details of Poincar\'e recurrences,  `virtual' thermal AdS phases do  control the overall noise amplitude  even at high temperatures where 
the thermal ensemble is dominated by large AdS black holes.  We also study  EPR correlations and find that, in contrast to the behavior of large correlation peaks, their noise level is the same in TFD states and in more 
general highly entangled states.

\vskip 0.5cm

\Date{April  2014}

\vfill

\vskip 0.1cm




\baselineskip=15pt

\newsec{Introduction}

\noindent

The very long time behavior of quantum systems  can exhibit peculiar properties. In particular,  
under certain conditions  Humpty Dumpty can  come back together again and Boltzmann brains can form.
The caution in attributing too much `practical' importance to these phenomena is rooted in the almost prohibitively long time scales
involved. One may well be concerned that many other yet to be discovered phenomena will show up well before these time scales are achieved, 
and they overshadow and modify the issues involved. 
These very long time  phenomena translate into very small effects in the  oscillatory behavior of correlation functions when averaged over time:  the quantum noise. 
Nevertheless these effects have been very useful in punching a hole in the arguments that information must be lost in the presence of
black holes and in testing the consistency of the AdS/CFT correspondence \refs\malda.

These issues have been discussed in some detail in the past \refs\us, and we return to them now, without waiting for a Poincar\'e time to elapse. 
The reasons for this study are two-fold. First it has been realized that fine structures in the noise can be analyzed within standard dynamical hypothesis from the theory of quantum chaos.  
Second, the suggestion that EPR=ER and its prequels (cf. \refs{\susmal, \malda, \mvr}) raised the question of whether one can use the properties of boundary CFTs to diagnose the association of a general 
entangled state somehow to geometric connectivity.
A point to appreciate is that the characteristics of the long time behavior of the correlation functions depends on many dynamical  ingredients  of the system,
the nature of the observables whose properties are probed and the particular states in which the system is placed. 

We begin in section 2 with an operational definition of quantum correlation noise in various types of correlation functions, including EPR systems.  
In section 3 we discuss the most suitable observables to probe the system once the dynamics is given, with a focus on constraints derived from the theory of quantum chaos. This leads to a choice of operators characterized by the Eigenvalue Thermalization Hypothesis (ETH). 
In section 4 we turn to estimate the value of the quantum noise and its dependence on the dynamics, the observables and the states. We 
find a diversity of relations between the value of the noise and the entropy of the system.
In section 5 we apply these results to AdS/CFT systems containing both black holes and graviton gases in their dual dynamical description. 
Finally, we conclude offering some speculations on the interpretation of the EPR=ER conjecture in the light of our results.

\newsec{ Quantum Noise Of Temporal Correlation Functions}

\noindent

In our discussion we will use the term `quantum noise' to describe the characteristics  of correlation functions after a certain time; essentially the time that is required for them to start oscillating around the long time average value they are supposed to attain. 
 To appreciate how this comes about consider first the example
of a  time 
self-correlation for an Hermitian operator $B$ in an infrared-bounded, unitary quantum system 
\eqn\bco{
C(t)  =\Tr\left[\,\rho\;B(t)\,B(0)\right]= \Tr \left[\,\rho \;e^{itH}\; B \;e^{-itH}\; B \; \right]
\;.}
The correlation \bco\ has a spectral representation
\eqn\spectral{
C(t) = \sum_{mn} \rho_{m}\,B_{mn}\,B_{nm} \,e^{i(E_m - E_n) t}\;,}
where we have taken for simplicity a diagonal density matrix in the energy basis. A particular case is provided by the canonical thermal state, $\rho_n = e^{-\beta E_n} / \sum_k e^{-\beta E_k}$, although the present discussion goes through for a general choice of $\rho_n$.  

The bounded character of the system is reflected in the discreteness of the energy spectrum. Indeed in this case the correlation is maximal at $t=0$ and decays away from there for a time period to be estimated.
For times small compared to the inverse level separation $\bra E_n - E_m \ket \,t \ll 1$ we may be able to approximate the discrete sums in \spectral\ by continuous integrals, resulting in either a power law or exponential decay,  depending on the detailed energy dependence of the operator matrix elements. However, the discreteness of the spectrum cannot be ignored over very long time scales, since the infinite time average of \spectral\ is strictly positive, \foot{We assume all spectral sums to be sufficiently convergent so that formal manipulations involving commutation of integrals and sums are permitted.}
\eqn\ita{
{\overline{C(t)}}\equiv \lim_{\tau\to\infty} {1\over \tau}\int_0^\tau dt \;C(t) = \sum_m \rho_m |B_{mm}|^2\;.
}
In fact,  the value of the limiting time average is controlled by the diagonal matrix elements of $B$ in the energy basis. The same diagonal matrix elements control the one-point  expectation value $\Tr [\rho B ]$. It will be convenient to work with operators having no
expectation values in energy eigenstates. In the following we will arrange for that by dealing with  modified operators $B$ from which the diagonal elements were subtracted.
 In such a situation the time average of the correlation vanishes  and we can define the noise from the time average of the modulus squared, 
 \eqn\noisedef{
 |{\rm noise}| \equiv \left[{\overline{|C(t)|^2}}\right]^{1/2} \;,}
 where 
\eqn\abs{
 {\overline{|C(t) |^2}} = \sum_{mnrs} \rho_m \rho_r |B_{mn} |^2 \,|B_{rs}|^2 \,\overline{e^{i(E_m -E_n + E_s -E_r)t}}
\;.}
When the spectrum is  generic,  that is  when no  rational relations exist between energy levels, the time average of the phase elements in \abs\  vanishes unless $E_m = E_r$ and $E_n =E_s$ or $E_m =E_n$ and $E_r =E_s$. The second option is discarded because we now assume $B_{nn} =0$. We are thus left with an average  noise amplitude
\eqn\absa{
|{\rm noise}|  = \left[ \sum_{mn} \rho_m^2 \,|B_{mn}|^4\right]^{1/2}\;,
}
which measures the average fluctuation amplitude of the function $C(t)$.

In general,  any two-point function has the structure  
\eqn\compos{
C(t) = \sum_{mn} C_{mn} (t) = \sum_{mn}  C_{mn} \,e^{i(E_n -E_m)t}
\;,}
with $C_{mn}$  depending quadratically  on the operator matrix elements. The component Fourier periods 
range from the shortest time scales  $t_{\rm short} \sim (\Delta E)_{\rm max}^{-1}$, determined by the largest energy difference in the 
the matrix elements of $B$, up to the longest time scales  $ (\Delta E)_{\rm min}^{-1} $  determined by the smallest energy  differences. The oscillatory structure is actually dominated by  the Heisenberg time scale $t_{\rm H} \sim ((\Delta E)_{\rm average})^{-1}$, which characterizes the average energy differences. 

On time scales in excess of the Heisenberg time most of the oscillation components in \compos\ have updated their phases by an amount of $O(1)$. In the absence of rational relations between the energy differences, the erratic behavior of $e^{i(E_m -E_n)t}$ will eventually erase any phase correlations that could exist at $t=0$ among the $C_{mn}$ coefficients, such as their uniform positive sign for the example at hand, $C_{mn} = \rho_m |B_{mn}|^2$. This means that, on time scales larger than $t_{\rm H}$,  all $C_{mn} (t)$ 
 can be regarded as having  uncorrelated phases. In this case the noise can be estimated by viewing each term in \compos\ as a step in random walk, i.e. by the root-mean-square $(\sum_{mn} |C_{mn}|^2)^{1/2}$. This heuristic method of analysis captures the properties that result  from the more formal definition  \noisedef. 

 A useful measure of the noise amplitude is obtained by comparison with the peak value, occurring at $t=0$ for the particular case of the correlation \bco:
 \eqn\max{
 |{\rm peak}| \sim|C(t)|_{\rm max} = \sum_{mn} \rho_m \,|B_{mn}|^2\;. }
 We then define  the ratio 
\eqn\normalnoise{
{|{\rm noise} |\over |{\rm peak}|}   = \left[{\sum_{mn} \rho_m^2 \, |B_{mn}|^4 \over \left(\sum_{mn} \rho_m \,|B_{mn}|^2 \right)^2}\right]^{1/2}\;.
}
Very roughly, there are two more  index sums in the denominator than in the numerator, so we expect \normalnoise\ to scale as $1/\CN_{\rm eff}$, where $\CN_{\rm eff} $ is some measure of  the number of states that are efficiently excited by the operator.  Following usual practice, we can define an effective `entropy' associated to this Hilbert-space dimensionality by the  formula $S_{\rm eff} = \log\,(\CN_{\rm eff})$, so that noise levels are expected to amount a fraction of order $e^{-S_{\rm eff}}$ of the peak value. 

A more detailed estimate  of $S_{\rm eff}$  requires
  some knowledge of the spectral properties of   the system,  the  operator and the particular quantum state. At any rate, if the long-time decay from the peak  is of approximate quasinormal type, proportional to $\exp(-\Gamma t)$, the correlation drops down to the  noise level after times of order
  \eqn\noiselevel{
  t_{\rm noise} \sim \Gamma^{-1} \,\log \left({|{\rm peak}| \over |{\rm noise}|}\right) \sim \Gamma^{-1} \,S_{\rm eff} 
  \;.
  }
  A milder, power-like decay proportional to $(\Gamma t)^{-\gamma}$, such as the one encountered for thermal diffusion of conserved charges, gives a characteristic relaxation time to the noise level of order $t_{\rm noise} \sim \Gamma^{-1} \, \left({|{\rm peak}| \over |{\rm noise}|}\right)^{1/\gamma}$. 
  
 Another general property of two-point correlations of the form \compos\ is their quasi-periodicity over extremely long time scales, associated to the general phenomenon of quantum Poincar\'e recurrences.  Following the previous discussion, let us assume that  \compos\ is well approximated by a subset of $O(\CN_{\rm eff}^2)$ terms
 in the sum over the $(m,n)$ indices. The configuration of all  time-dependent phases is determined at any time by that of the $\CN_{\rm eff}$ phases $e^{iE_n t}$, which represent a point on the torus ${\bf T}^{\CN_{\rm eff}}$. Since the energies are assumed to have no rational relations, time evolution induces an ergodic motion on this torus, so that any region of volume $v$ is visited in a time proportional to $1/v$. With accuracy  $\Delta \alpha$ for each of the $e^{iE_n t}$ phases we have $v=  (\Delta \alpha /2\pi)^{\CN_{\rm eff}}$ and the return time
is of order
\eqn\deltaal{
t_{\Delta \alpha} \sim \bra E\ket^{-1} \,\exp\left(\CN_{\rm eff} \,\log (2\pi /\Delta\alpha)\right)
\;,}
where the $\bra E\ket$ is the average energy, setting the average velocity of the $\CN_{\rm eff}$ `clocks'  undergoing ergodic motion. If we are just interested in the return  of a configuration within  $O(1)$ accuracy, we can take $\log (2\pi /\Delta\alpha) =1$ and find the so-called quantum Poincar\'e time  $t_{\rm P} \sim \bra E \ket^{-1} \exp(e^{S_{\rm eff}})$ scaling as a double exponential of the effective entropy.

 \bigskip
\centerline{\epsfxsize=0.6\hsize\epsfbox{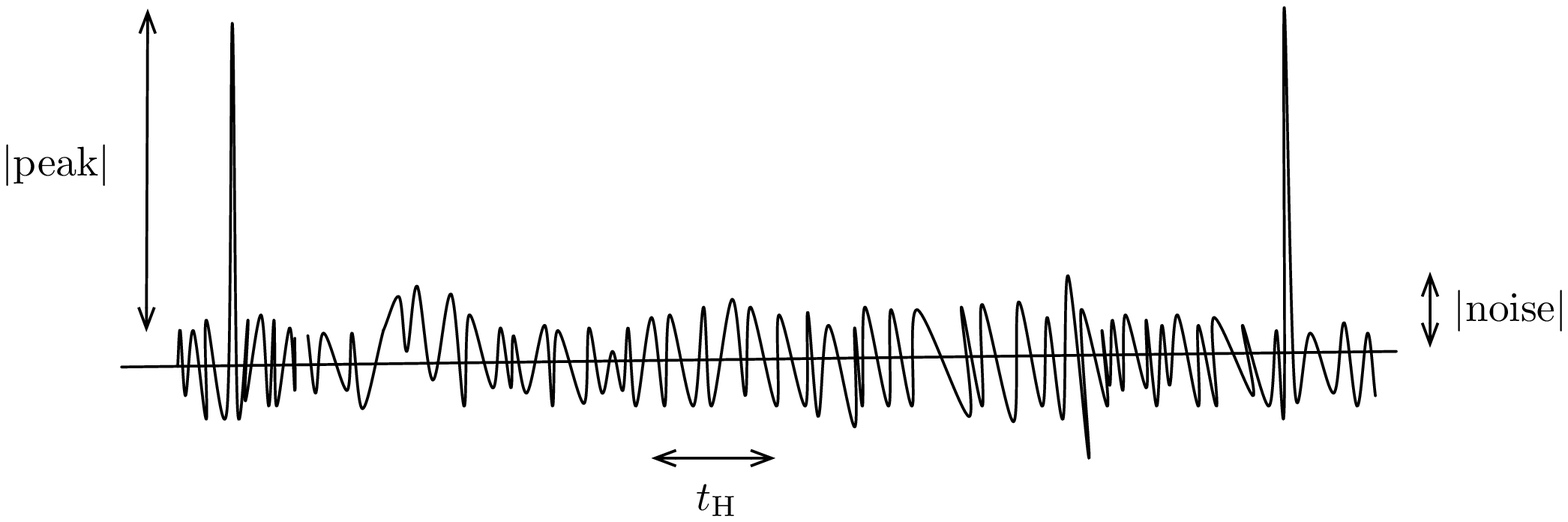}}
\noindent{\ninepoint\sl \baselineskip=2pt {\bf Figure 1:} {\ninerm
A rendering of the quantum oscillation noise in the time graph of a correlation function. The $O(1)$ correlation peaks are separated by Poincar\'e times $t_{\rm P}$. The basic oscillation period is the Heisenberg time $t_{\rm H}$, although many faster wiggles coexist in building the noise  fluctuations of  $O(|{\rm noise}|)$. The time width of the large peaks is the noise relaxation time $t_{\rm noise}$. In chaotic systems these three time scales are often exponentially related to one another, i.e. $t_{\rm H} \sim \log (t_{\rm P})$ and $t_{\rm noise} \sim \log (t_{\rm H}) \sim S_{\rm eff}$ in appropriate units. The scrambling time \refs\scr, too small to fit the figure, is also conjectured to give a further step in the exponential hierarchy: $t_{\rm scram} \sim \log (t_{\rm noise})$.
}}
\bigskip

The different time scales introduced  in this section (cf. figure 1)  depend on dynamics, choice of correlation function and choice of state  in different ways. It is the description of the different setups which will occupy a large part of this work. Along the way we will update and make more precise  some of the estimates made previously  in \refs{\oldsus, \us}.

   \subsec{More General Probes}
  
  \noindent
  
  The thermal self-correlation considered in \bco\ is an example of a correlation with a clearly identifiable maximum value at $t=0$. One can consider more general two-point correlations with different operators and/or states of the form
  \eqn\cene{
  G_{BB'} (t, t') = \Tr\left[\,\rho\, B(t)\,B' (t')\,\right] = \sum_{mnr} \,\rho_{mn} \, B_{nr}\,B'_{rm} \,e^{i(E_n - E_r)(t-t')} \,e^{-i(E_m -E_n)t'}
  \;,
  } 
  where $\rho_{mn}$ is a general density matrix characterizing the state of the system. 
  If the state (density matrix) is stationary, i.e. diagonal in the energy basis, the correlation only depends on $t-t'$ and we can
  set $t'=0$ with no loss of generality. For general non-stationary density matrices we can rewrite \cene\ as
  \eqn\oooo{
  G_{BB'}(t,t') =\Tr\left[\,\rho(t')\,B(t-t') \,B(0)\,\right] \;,}
  with $\rho(t') = e^{-it' H} \rho\, e^{it'H}$, and ascribe the $t'$ dependence to the specification  of the state. If we are interested in generic properties, we may put $t'=0$ and absorb the $t'$ dependence on the generic choice of $\rho$.

   We shall denote correlations in stationary states by $D_{BB'} (t)$, to signify the diagonal character of the density matrix. 
  In general,  \cene\ does not peak at $t=t'$, even for $[\rho, H] =0$. This would  require conditions  on the matrix elements of $B$ and $B'$, i.e. a concrete {\it correlation} between the operators, which in general will be state-dependent.
  
  In many situations it is interesting to consider the doubled version of the system to purify the mixed state $\rho$. We refer to the original system as `Bob' and to the purification copy as `Alice'. A general  normalized state of the form
  \eqn\gens{
  |G\ket = \sum_{mn} g_{mn} \,|m\ket_A \otimes |n\ket_B 
  \;,
  }
  defines a density matrix on  Bob's side 
  \eqn\gob{
  \rho_{nn'} = \sum_m g_{mn} \,g^*_{mn'}
  \;,}
   so that \cene\ can be obtained as a $G$-expectation value of Bob-side operators $B$ and $B'$. 
   
   More generally, we can consider the so-called EPR correlations between
`Alice' and `Bob' operators 
\eqn\eprc{
G_{AB} (t_A, t_B) = \bra A(t_A) \,B(t_B) \ket_{G} = \bra G | e^{-it_A H} \, A\, e^{it_A H} \,e^{it_B H} \,B\,e^{-it_B H} | G\ket\;,}
where we adopt the common convention of inverted time flow on the Alice side. With this definition \eprc\ has similar time-dependence  properties as \cene, since we can rewrite \eprc\ as
\eqn\eprco{
G_{AB}(t_A, t_B) = \bra G(t_B) \,|\,A(t_A -t_B)\,B(0)\,|\,G(t_B)\ket\;,
}
where 
$$
|G(t)\ket \equiv \sum_{mn} g_{mn}\,e^{it H} |m\ket_A \otimes e^{-itH} |n\ket_B\;.
$$
States with diagonal entanglement in the energy basis, $g_{mn}  \propto \delta_{mn}$, are stationary and the corresponding EPR correlation \eprco\ only depends on $t_A -t_B$. In this stationary situation  we can set $t_A -t_B =t$ and denote the correlation as $D_{AB} (t)$. 
  One such example is the `thermo-field double' state (TFD) with $g_{mn} = Z(\beta)^{-1/2} \,e^{-\beta E_n /2} \delta_{mn}$, whose `one-sided' density matrix is the standard canonical ensemble $\rho_T$.   If we are interested in properties  of  \eprco\  for generic choices of $g_{mn}$ we can also absorb the $t_B$ dependence in the choice of state $|G\ket$ and set $t=t_A -t_B$. 

EPR correlations can always be pulled back to `one-side' correlations if the detailed double state is known. This is achieved by
diagonalizing the entanglement in the so-called Schmidt basis:
\eqn\sch{
g_{mn} = \sum_\alpha (\Omega_A)_{m\alpha} \,\sqrt{\rho_\alpha} \, (\Omega_B)_{n\alpha}\;,
}
where the notation reflects our ability to define the `entanglement eigenvalues' $\sqrt{\rho_\alpha}$ as positive definite by absorbing phases in the independent unitary matrices $\Omega_A$ and $\Omega_B$. The $\rho_\alpha$ are also the eigenvalues of the Bob-side density matrix,\foot{Notice that Bob and Alice can have different density matrices in the energy basis, but both have the same eigenvalues.} so that they satisfy $\sum_\alpha \rho_\alpha =1$,
\eqn\dens{
\rho_{nn'} = \sum_\alpha \rho_\alpha \,(\Omega_B)_{n\alpha} \,(\Omega_B)^*_{n'\alpha}\;.}

In the case of doubled EPR states, the $\rho_\alpha$ measure the degree of entanglement, ranging from zero in the case
that only one $\rho_\alpha$ is non-zero, to maximal entanglement when all of them are equal to one another. The matrices $\Omega_A$ and $\Omega_B$ measure the departure from `diagonal' entanglement, by which we refer to the alignment between the Schmidt basis which diagonalizes entanglement and the energy basis which diagonalizes the Hamiltonian. On Bob's side, `alignment' is equivalent  to $[\rho, H]=0$, i.e. stationarity of the Bob-side state.

We can rewrite \cene\ in the Schmidt basis as
\eqn\cenee{
\Tr\left[\rho\,B(t)\,B'(t')\right] = \sum_{\alpha\beta} \rho_\alpha \left(\Omega_B^\dagger\,B(t)\,\Omega_B\right)_{\alpha\beta} \left(\Omega_B^\dagger \,B'(t')\,\Omega_B \right)_{\beta \alpha}\;,
}
while the EPR correlation \eprc\ takes the form 
\eqn\eprnn{
\bra A(t_A) \,B(t_B)\ket_G = \sum_{\alpha\beta} \sqrt{\rho_\alpha \rho_\beta} \,\left(\Omega_A^\dagger \,A(t_A) \,\Omega_A \right)_{\alpha\beta} \left(\Omega_B^\dagger \,B(t_B)\,\Omega_B\right)_{\alpha\beta}\;. 
}
If the $G$-state has sufficient entanglement so that all $\rho_\alpha \neq 0$, knowledge of  $\Omega_A$ can be used  to construct a surrogate of Alice on Bob's side which gives the same EPR correlation with the rule
\eqn\rule{
\bra \,A(t_A) \,B(t_B)\,\ket_G = \Tr\left[\rho\,B(t_B) \,B_{A(t_A)}\right]\;,
}
with
\eqn\surrog{
\left(B_{A(t_A)}\right)_{\alpha\beta} = \sqrt{\rho_\alpha}\, \left(\Omega_A^\dagger\, A(t_A)\,\Omega_A\right)_{\beta\alpha} \,{1\over \sqrt{\rho_\beta}} \;. 
}
A particular case of this relation is well known for the case of the TFD state, where  it can be written as an analyticity property:
\eqn\tfdm{
\bra A (t_A) B(t_B) \ket_{\rm TFD} = \Tr \left[ \rho_{T} \,{\widetilde A}(t_A -i\beta/2) B(t_B) \right] \;,
}
where ${\widetilde A}_{mn} = A_{nm}$ and the time evolution in the r.h.s. is defined with Bob's time arrow. This convention is different from the one implicit in \surrog, where the time evolution of $A(t_A)$  is performed on the Alice side and the result is mapped to Bob's side with \surrog. 

The analyticity relation \tfdm\ is very important in AdS/CFT calculus, since all correlation functions computed via bulk rules can be
obtained as analytic continuations of Euclidean bulk correlations on the Euclidean bulk saddle point manifolds, such as the eternal black hole `cigar' metric. This means that the bulk computational rules are implicitly designed for a large-$N$ limit of  TFD states (or appropriate deformations thereof.) It is for these states for which there is a strong case for the EPR=ER conjecture of \refs\susmal, while the `geometric' status of more general highly entangled states  is still under study \refs{\polmarol, \shenstan, \susnew}. 

\newsec{Representative Dynamics And Observables}

\noindent

 Noise levels in correlations depend on detailed dynamical properties regarding both the spectrum of the Hamiltonian and the particular choice of quantum state and operators. 
With a very broad brush we can distinguish two extreme cases. One extreme case occurs when the system is close to being integrable, like a gas of almost free quasiparticles. In this familiar situation it is natural to choose $B$ as a one-particle operator, whose  matrix elements  in the energy basis are very sparse, connecting only states differing by one unit of quasiparticle number. 

The other extreme is a more generic non-integrable system with chaotic dynamics in the quantum sense. A practical criterion for quantum chaos has been proposed in   \refs{\peres, \deutsch, \srednicki} along the following lines:  given a generic observable $B$ which does not commute with the Hamiltonian, we have chaotic dynamics when the eigenstates of $B$ can be considered as  uncorrelated  with the eigenstates of  $H$, so that both basis are related by a random unitary transformation. This suggests  that those $B$ eigenstates will be very efficiently mixed under the time evolution generated by $H$. 

More formally,  let us restrict $B$ to the subspace generated by a narrow band with  $e^S$ energy eigenstates. We say the band is   `chaotic' with respect to the observable $B$ if  the unitary matrix $U$ which diagonalizes it, 
$$
B_{mn} = (U\, b \, U^\dagger)_{mn} = \sum_\alpha b_\alpha \,U_{m\alpha}\,( U_{n\alpha})^*\;, 
$$
looks like a  random unitary matrix.\foot{Perhaps more properly, we would characterize $U$ as {\it pseudorandom}, since both $B$ and $H$ are fixed from the beginning.} In this expression, $b$ denotes the diagonal matrix of $B$ eigenvalues. From $\sum_{\alpha} |U_{\alpha n}|^2 =1 $ we learn that the matrix elements of $U$ have typical size $e^{-S/2}$. If the $b_\alpha$ are not very special (such as  $O(e^S)$ of them being equal) the off-diagonal entries of $B_{mn}$ are sums of $e^S$ terms with random phases and  size $e^{-S}$ each, resulting in an overall estimated size of  $O(e^{-S/2})$. Diagonal entries are of the form $B_{mm} = \sum_\alpha b_\alpha |U_{\alpha m}|^2$ which scales like $e^{-S} \Tr B$.  Hence, the diagonal elements of $B$ within the chaotic band are measured by the average size of $B$'s eigenvalues within the band. Notice that these diagonal elements may end up being of $O(1)$  if the  $b_\alpha$ have mostly the same sign. 

The arguments above have led us to what is sometimes called  the Eigenvalue Thermalization Hypothesis (ETH), which is an ansatz  for the statistical properties of  $B_{mn} = \bra E_m | B | E_n \ket$ as we vary the $(m, n)$ indices over their $e^{2S}$ values. Namely $B_{mn}$  has typical size $e^{-S/2}$ with erratic, random-walking phases, except for the diagonal terms whose size is controlled by the detailed properties of the $B$ eigenvalues. 

This rule applies only to those energy matrix elements of $B$  connecting states   within the chaotic band. No specific statement is made regarding other off-band matrix elements but,  for the purposes of  applying estimation techniques, it is useful to  simplify matters by restricting attention to smeared or regularized operators, whose energy width $\Gamma_B$  is cut-off to coincide with the   width of the chaotic band itself, $\Delta$. 
On the other hand, the band is considered `narrow' when the density of states does not vary significantly over the interval of width $\Delta$. An $O(1)$ multiplicative variation  of the density of states corresponds to an $O(1)$ additive variation of the entropy. Using the general rule $\Delta S \approx \Delta E /T(E)$, with $T(E) = (\pt S / \pt E)^{-1}$ being the microcanonical temperature, we can conventionally define an energy band as `narrow' when $\Delta \sim T$. Combining this condition with the rule for the widths of the class of operators we are considering, we conclude that the
chaotic form of matrix elements should hold for operators with $\Gamma_B \sim T$, i.e. for operators smeared over the effective thermal length of the state. 

Even if an operator with $\Gamma_B \gg T$ can be defined in principle, it is natural to assume that, after a time of the order of the `scrambling time',  the behavior of the correlations is  dominated by the matrix elements with  $|E_n - E_m | \sim T$. For example, a single-trace operator in AdS/CFT, acting on a large black hole state, injects a particle from the boundary of AdS with an energy which can be much larger than $T$, but this state falls to the Rindler region in a time of order $\beta=1/T$, and even further to the stretched horizon  in a time of order $\beta \log (N^2)$, where it is expected to scramble in a time of the same order of magnitude (cf. \refs{\scr,\javi}). In this paper we are primarily concerned with time scales much larger than the typical scrambling time scales, and thus we shall perform estimates under  the simplifying assumption  that operators on chaotic bands are sharply  cutoff to an energy width $\Gamma_B \sim T$. 

Considering a smooth spectral modulation governed by an interpolating   entropy function  $S(E)$, we have the continuous generalization of ETH \refs\srednicki: 
   \eqn\eth{
 B_{mn}  = {\bar B} ({\bar E}) \delta_{mn} +b({\bar E}, \omega) \,e^{-S({\bar E})/2} \,R_{mn}\;,
 \qquad 
  {\bar E}=\shalf(E_m + E_n)\;,\;\;\;\omega = E_m - E_n\;.
 }
 The function ${\bar B}({\bar E}) $, being a coarse-grained measure of the eigenvalue's average, is smooth in ${\bar E}$. Similarly   $b({\bar E}, \omega)$ measures the energy width in the off-diagonal directions. It has a characteristic support in $\omega$ of order    $\Gamma_B ({\bar E}) $ and it is also smooth in its arguments, since the erratic component of the matrix elements is already parametrized by 
 $R_{mn}$, a  matrix with entries of $O(1)$ and erratic phases as a function of $(m,n)$ (cf. figure 2). A crucial property of the ETH ansatz is its stability under products: if $A$ and $B$ satisfy ETH, then $AB$ also satisfies ETH.   This  fact is the key to ETH  being useful in proving the  basic tenets of statistical thermodynamics, such as ergodicity and mixing on approach to equilibrium. 
 
 \bigskip
\centerline{\epsfxsize=0.6\hsize\epsfbox{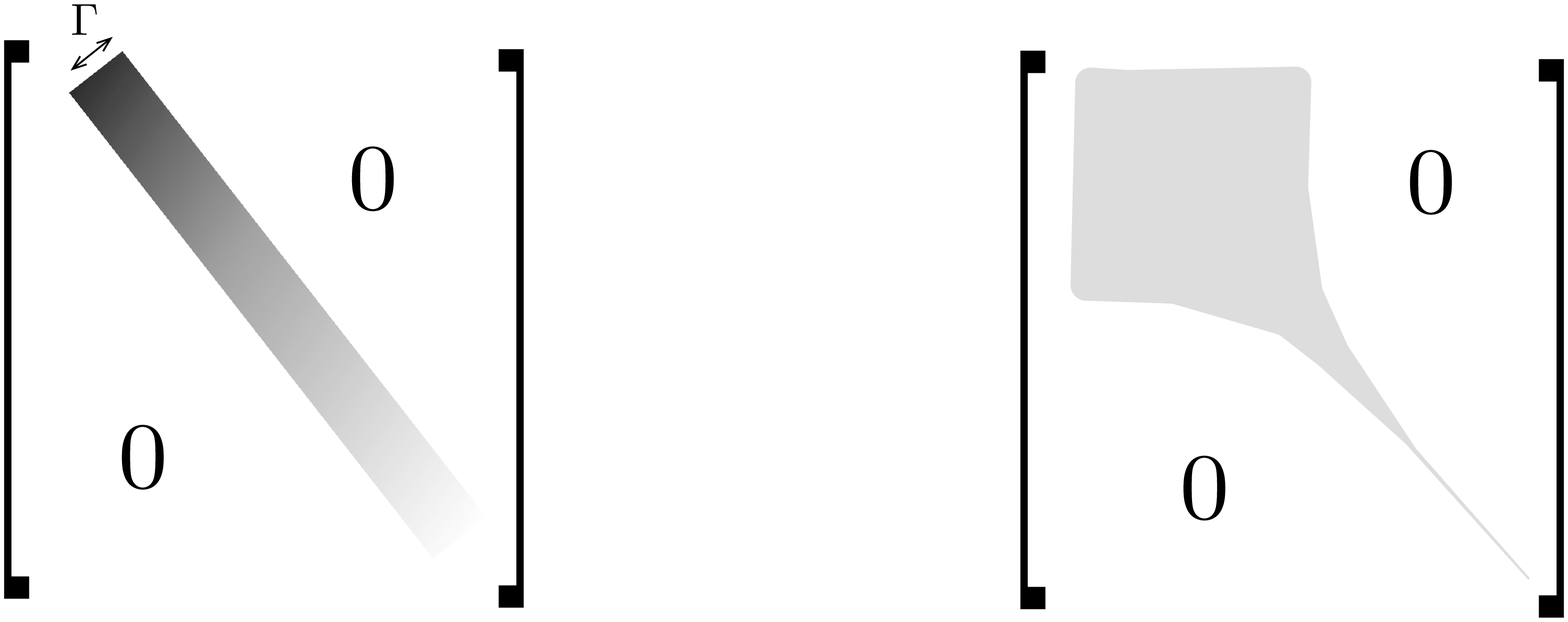}}
\noindent{\ninepoint\sl \baselineskip=2pt {\bf Figure 2:} {\ninerm
On the left, the graphical representation of a regularized ETH operator with energy width $\Gamma$ on both high and low energy bands. The entries are quasi-continuous, but their density is much larger (dark shading) in the high energy region (upper left  corner). On the right, the same operator written with uniform-density matrix-element notation, where now the width represents the number of entries rather than their energy separation. 
}}
\bigskip

 In the application to AdS/CFT one encounters natural representatives of both extreme models for the dynamics. CFTs with a smooth gravity dual have specific spectral properties. In particular,  the high-energy entropy scales as  the central charge, of order $N^2$,  dominated by states which are described as black holes in the bulk picture (plasma of glue in the CFT picture.) In addition, there is a band of low-energy states looking like graviton-gas states in the bulk  (glueballs in the CFT) with entropy of $O(1)$ in the large $N$ limit. The low energy band extends from the spectral mass gap up to the energies of $O(N^2)$ where the `black hole' states start to dominate the density of states.\foot{Fine details of the spectrum may include additional intermediate bands with different types of black holes and/or Hagedorn transients.} The natural dynamical assumption is that of quantum chaos for the high-energy (black hole) band and thermal gas for the low-energy (graviton) band. Interactions among gravitons are suppressed by powers of $1/N^2$. Hence, at moderate values of  $N$ we may  consider also a chaotic model for an interacting graviton gas. In this case the main difference between the high and low energy bands would be just the jump in density of states.

With the AdS/CFT application in mind, special simplifications occur when restricting attention to regularized operators whose fixed energy width  $\Gamma_B$ is small compared to the overall energy range of interest. For example, we may consider a model with two bands (high and low) with $E_h - E_l \gg \Gamma_B$. In this case we can approximate the 
 $B_{mn}$ matrix by a block-diagonal form as in figure 3. 

\bigskip
\centerline{\epsfxsize=0.3\hsize\epsfbox{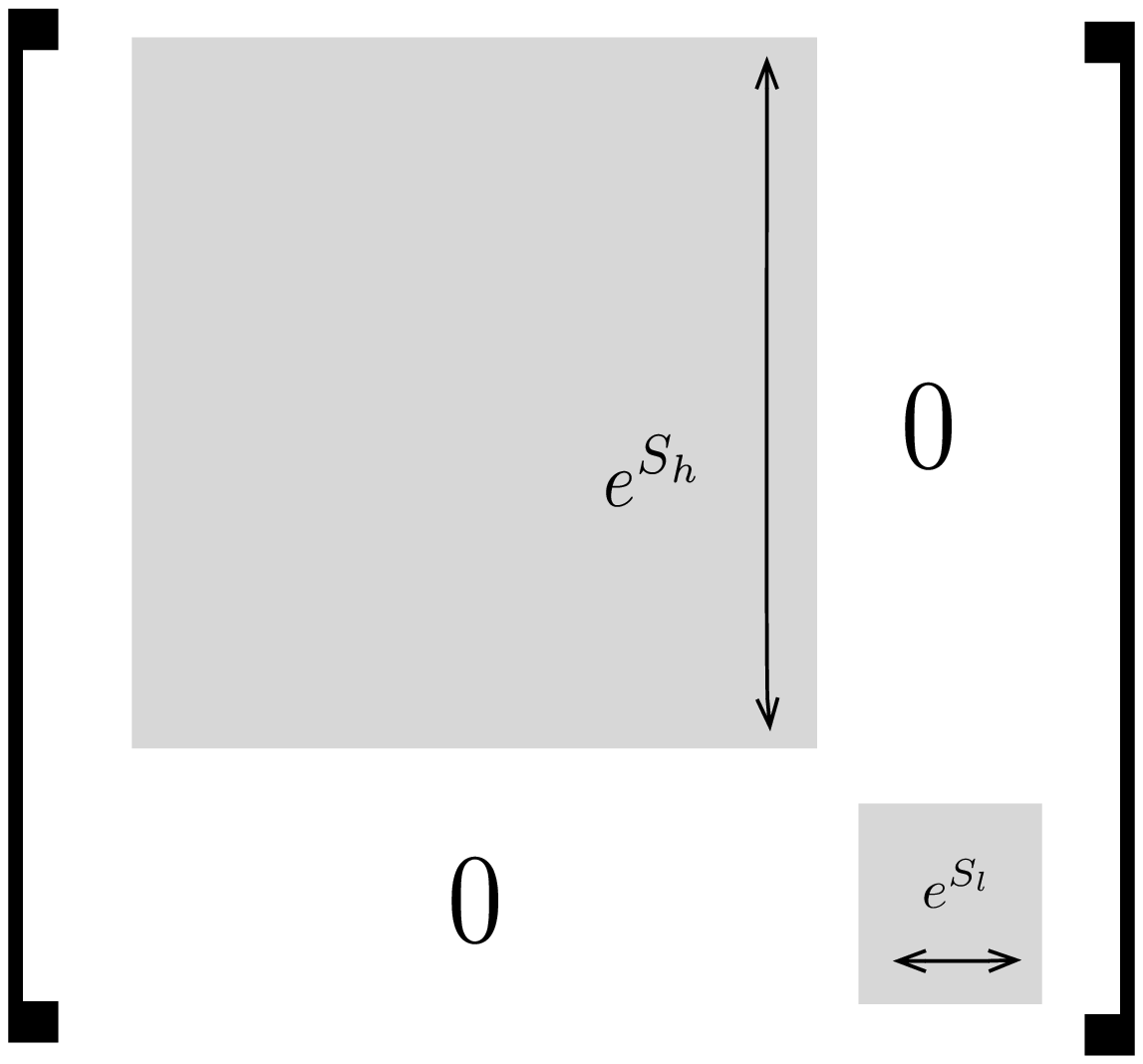}}
\noindent{\ninepoint\sl \baselineskip=2pt {\bf Figure 3:} {\ninerm
If the relevant part of the spectrum is approximated by two narrow bands with very different average energies, $E_h \gg E_l$, and dimensionalities,  $e^{S_h} \gg e^{S_l}$, the ETH ansatz for regularized operators can be simplified by writing a block-diagonal form as in the figure, in the discrete uniform density representation of matrix entries. If the upper band is chaotic and the  lower band has perturbative quasiparticle-like states, a single-particle operator would have the ETH form in the upper band and the sparse quasiparticle form in the lower band.
}}
\bigskip

\subsec{Correlations And ETH}

\noindent

 As argued in section 2, the discussion of  the average noise amplitude focuses on time scales in excess of the Heisenberg time $t_{\rm H}$ and thus is hardly sensitive to the occurrence of correlation peaks.
However, it is important to determine the refinements of the ETH ansatz that are implied by their presence.  Any such correlation peak requires a degree of phase correlation between the corresponding operators. A simple example is provided by the  self-correlation  \bco, where the  product of matrix elements  $B(t)_{mn} B(0)_{nm}$ equals $|B_{mn}|^2$ at $t=0$. It is the sign coherence at $t=0$ what builds a peak in the correlation function.  More generally, for distinct operators $B$ and $B'$, we may refine the ETH ansatz by requiring a phase correlation (cf. \refs\polmarol) 
  \eqn\corp{
 (R^B)_{mn} \, (R^{B'})_{rs} = (\CD_{BB'})_{mn} \,\delta_{ms} \, \delta_{nr} + ({\rm erratic})_{mnrs} \;,
 }
 with $(\CD_{BB'})_{mn}$ having `smooth' phase orientation as a function of the $(m,n)$ indices. There is an inherent ambiguity in the splitting between `smooth' and `erratic' terms, as the notion of smoothness for a discrete function only becomes well defined in the continuum limit. We will deal with this continuum limit shortly, but for now we can say that $(\CD_{BB'})_{mn}$ is `smooth' when it can be approximated by an index-independent matrix over a sufficiently narrow energy band. The erratic component represents the fluctuations around the correlated term and it is of the same order of magnitude. 
 
 The phase correlations imposed by the first term in \corp\   induce
 an $O(1)$ peak  at $t=0$ in any Bob-side correlation function with stationary density matrix: 
 \eqn\statdenc{
 D_{BB'} (t)= \sum_{m n} \rho_m \,B_{mn} \,B'_{nm} \,e^{i(E_m - E_n)t}\;.}
 To see this, we pick the first term in \corp\ to find the `smooth' component of the correlation as 
 \eqn\smodenc{
 D_{BB'}^{(s)} (t) =  \sum_{mn} \rho_m \, \left(b\,b'\,\CD_{BB'} \,e^{-S}\right)_{mn} \,e^{i(E_m - E_n)t} \;,
 }
 which is of $O(1)$  at $t=0$ as required,  as a result of the smoothness of the $(m,n)$ index dependence.  On time scales in excess of the Heisenberg time of the system \smodenc\ gives a contribution to the fluctuation noise of $D_{BB'} (t)$, but we will see in the next section that the noise is always dominated by the erratic component in \corp.  On the other hand, on time scales much smaller than the Heisenberg time the time-dependent phases in \smodenc\ vary slowly, and we can approximate the index sums by integrals to give a parametrization of the descent from the peak at $t=0$.

 The integration measure can be defined with an interpolating entropy function $S(E)  = \log n(E)$, where $n(E)$ is a smooth function measuring the number of energy levels up to energy $E$, which leads to 
 \eqn\conm{
 \sum_{E_n} f(E_n) \approx \int dE  {dn \over dE} \,f(E) = \int dE \,\beta (E) \,e^{S(E)} \,f(E)\;,
 }
 for smooth functions $f(E_n)$ of the energy levels, 
 where  $\beta(E) = \partial S / \partial E$ is 
 the microcanonical inverse temperature. Using this prescription in combination with the smooth form of the ETH ansatz \eth\ we can write 
 \eqn\smoocos{
 D_{BB'}^{(s)} (t) \approx \int dE_m \,\beta(E_m) \,e^{S(E_m)} \,dE_n \,\beta(E_n) \,e^{S(E_n)} \,\rho(E_m) \,F_{BB'} ({\bar E}, \omega) e^{-S({\bar E}) }\,e^{i\omega t}\;,}
 where we use ${\bar E} = \shalf (E_m + E_n)$,  $\omega = E_m -E_n$ and denote 
 $$
 F_{BB'} ({\bar E},\omega) = \left(b\,b'\,\CD_{BB'}\right)_{E_n\, E_m}\;.$$
 In view of the comments above, we can regard the smoothness of the $F({\bar E}, \omega)$ function as the operational definition of what we mean by first term on the right-hand side of \corp. 
 
 With the assumption of narrow width of order $T(E)$  for the operators, we can approximate the integrand by its value along the $\omega=0$ diagonal, up to corrections of relative order $T({\bar E}) / {\bar E} \sim 1/S({\bar E})$ to obtain
 \eqn\finsmoo{
 D_{BB'}^{(s)} (t) \sim \int d{\bar E} \beta({\bar E})^2 \, e^{S({\bar E})} \rho({\bar E}) \,\int d\omega \,F_{BB'} ({\bar E}, \omega) \,e^{i\omega t}\;.
 }
 In this expression, the integral over ${\bar E}$ gives the thermodynamic average and the integral over $\omega$ gives the
 time-dependence.  Assuming some standard form for the density matrix, such as the canonical $\rho(E) = Z(\beta)^{-1} e^{-\beta E}$,
 the ${\bar E}$ integral can be approximated by saddle point to obtain
 \eqn\smaa{
 D_{BB'}^{(s)} (t) \sim Z(\beta)^{-1} \sum_b \,e^{-I_b (\beta)} \int d\omega \,F_{BB'}^{(b)} (\omega) \,e^{i\omega t}\;,
 }
 where the index $b$ now denotes the different saddle points, $I_b (\beta) = \beta E_b - S(E_b)$  and $F_{BB'}^{(b)} (\omega) = F_{BB'} ({\bar E}_b , \omega)$.  
 
 If our quantum system admits a duality of AdS/CFT type,  it is natural to interpret 
 \smaa\ as the gravity dual version of the correlation with the index $b$ ranging over topologically different backgrounds of the bulk
 theory:
 \eqn\gravc{
 D_{BB'}(t)_{\rm bulk} \approx Z(\beta)^{-1} \sum_{X_b} e^{-I(X_b)} \bra \phi_B (t) \,\phi_{B'} (0)\ket_{X_b}\;,
 }
 where $Z(\beta) \approx \sum_{X_b} e^{-I(X_b)}$. This is the bulk semiclassical expansion as a sum over backgrounds $X_b$ compatible with the asymptotic boundary conditions of the canonical ensemble and  Euclidean action $I(X_b)$.  The bulk fields $\phi_B $ and $ \phi_{B'}$ entering the background-dependent two-point function are inserted on the boundary of $X_b$. This perturbative two-point function in each bulk  manifold is identified with
 \eqn\di{
 \bra \phi_B (t) \,\phi_{B'} (t)\ket_{X_b} \sim \int d\omega \,F_{BB'}^{(b)} (\omega) \,e^{i\omega t}
 \;. }
 For example, if 
 $F^{(b)} (\omega)$ has a set of poles $\omega_l$ with $|{\rm Im}\,\omega_l| = \Gamma_l$, we have all the ingredients to obtain
 the quasinormal long-time behavior  characteristic of black-hole backgrounds. 
 On the black-hole `cigar' background the Euclidean time translation isometries have a fixed point at the horizon. This implies a continuous spectrum for the real time Hamiltonian of perturbative excitations. The end result is the absence of noise component in all correlations obtained by bulk rules in `black' backgrounds (cf. \refs{\us, \more, \festucciaII}.) One typically finds quasinormal behavior for all correlations (either EPR or single side) as $t\rightarrow \infty$:
\eqn\qus{
D(t)_{\rm black} \approx \sum_l K_l  (t) \,e^{-\Gamma_l t} \;, }
with $K_l$ a set of oscillating functions depending on the operators and the dominant (minimum) $\Gamma_l $ damping coefficient of the order of the temperature $T$.  The form \qus\ ensures that the correlations vanish at large times and give no contribution to the infinite time average of \ita\ or \abs. It also shows no sign of the phenomenon of Poincar\'e recurrences expected in bounded systems.

 Hence, we conclude that the bulk computational rules of AdS/CFT can be interpreted in two steps. First we keep the `smooth' piece  \smodenc\ in the correlations of ETH operators.  This is necessary to capture the $O(1)$ peak in the correlation. Second, we further apply the  approximation of continuous energy spectrum,  while leaving out all the noise components as non-perturbative information. Notice that, while \smodenc\ has noise and Poincar\'e recurrences, these are no longer visible once we apply the   continuum-spectrum approximation around the peak \smoocos, and they are also absent from the gravity version \gravc. It is in the second step that we effectively decouple the Heisenberg and Poincar\'e  times. 
 
 An analogous refinement of the ETH ansatz is required to engineer peaks in EPR correlations for the doubled Alice-Bob system. In this case we require
 \eqn\corpa{
 (R^A)_{mn} \, (R^B)_{rs} = (\CD_{AB})_{mn} \,\delta_{mr} \,\delta_{ns} + ({\rm erratic})_{mnrs} \;,
 }
 with smooth $(\CD_{AB})_{mn}$, to have an $O(1)$ correlation peak for any EPR correlation function of type $D_{AB} (t)$, whose entanglement is diagonal in the energy basis. The discussion of the continuous approximation on descending from the peak
 is identical to the single-side one. In this case, the use of states with aligned entanglement ensures the analytic continuation
 \tfdm, which now translates into  analytic properties of the single-particle spectral function $F_{AB} (\omega)$ in the complex $\omega$
 plane (cf.  for example \refs\festucciaI)

 \newsec{Noise Estimates}
 
 \noindent
 
 In this section we will proceed to estimate the noise in the various systems we have now presented. Most of the discussion will
 deal with chaotic dynamics within the ETH ansatz expressed above, although we will briefly consider the main properties of quantum noise in systems of weakly coupled quasiparticles in the last subsection. 
 
 For pedagogical reasons, it is useful to discuss the more intricate issues of operator and state dependence in the simpler setting of a narrow energy band of average energy $E$, dimensionality $e^S$ and width $\Delta \sim E/S \sim T$, where $T$ is an average microcanonical temperature on the band. Chaotic operators are assumed  to satisfy ETH within this band, with energy cutoff of order $T$, and all smooth functions in 
 \eth\ are regarded as constant within the band, i.e. we take
 \eqn\ethband{
 B_{mn} \sim b\,e^{-S/2} \,(R^B)_{mn}\;,
 }
 with constant $b$. The requirement of vanishing diagonal elements can be obtained by the appropriate constraint on the erratic
 matrix $R^B$.  For chaotic dynamics, 
the Heisenberg time scale is of order $t_{\rm H} \sim e^S /\Delta$ and the Poincar\'e time scale is $t_{\rm P} \sim \exp(e^S)/E$.

 Within the refined ETH ansatz, all two-point functions  split into  {\it smooth} and {\it erratic}  components according to the decomposition in \corp\ and \corpa, with the $\CD_{mn}$ matrices regarded as an $O(1)$ constant independent of the indices, again as a result of the band narrowness. 
 
In what follows we estimate the noise level of general correlations $G_{BB'} (t)$ and $G_{AB} (t)$ for a choice of operators  as specified in \ethband,  \corp\ and \corpa. We find it interesting to track the noise contributions of both terms in \corp\ and \corpa,
despite the fact that the contribution to the noise of the `smooth' term is affected by possible ambiguities, and only the `erratic' piece is expected to give a reliable measure of the noise level. 
 
 \subsec{Bob's Noise}
 
 \noindent
 
 We begin with the general one-side correlation \cenee,  written in the form 
 \eqn\ones{
 G_{BB'} (t) = \sum_\alpha \rho_\alpha \left(\Omega_B^\dagger B(t) B' (0) \Omega_B \right)_{\alpha\alpha}
 \;.}
 Stationary states correspond to $\Omega_B ={\bf 1}$ and arbitrary non-stationary states  correspond to generic $\Omega_B$ with entries of size $e^{-S/2}$. Pure states have only one non-vanishing $\rho_\alpha$, equal to unity, while highly  mixed states have all $\rho_\alpha \sim e^{-S}$.

 The smooth correlation piece is obtained by selecting the first term in \corp. It has the form
 \eqn\peakgen{
 G_{BB'}^{(s)} (t) \sim|b\,b'\,| \, e^{-S} \sum_\alpha \rho_\alpha \, \sum_{mn} (\Omega_B^\dagger)_{\alpha m} \,e^{i(E_m - E_n)t} \,(\Omega_B)_{m\alpha}
 \;,}
 where we have approximated $b$, $b'$ and $\CD \sim O(1)$ as constant matrices over the narrow band. 
 
 At $t=0$ the $n$-sum cancels the $e^{-S}$ factor and the $m$ sum gives $(\Omega_B^\dagger \Omega_B)_{\alpha\alpha} =1$. Since
 $\sum_\alpha \rho_\alpha = \Tr \rho =1$, we recover the $O(1)$ peak at $t=0$,  independently of the value of $\Omega_B$, i.e.
 the peak exhibited by  the smooth piece of single-side correlations is independent of the stationary character  or the purity of the state. 
 
 To determine the asymptotic noise level of \peakgen\ we look at  large times. Then the $n$-sum over the ergodic phase   $e^{-iE_n t}$ gives an overall factor of $O(e^{S/2})$. Furthermore, the erratic phase $e^{iE_m t}$ destroys the coherence of the product of  $\Omega_B^\dagger $ and $\Omega_B$, 
 $$
 \sum_m (\Omega_B^\dagger)_{\alpha m}\, e^{iE_m t} \,(\Omega_B)_{m\alpha} = \sum_m e^{iE_m t} |(\Omega_B)_{m\alpha}|^2 
\;,$$
which is of order $e^{-S/2}$ with a random $\alpha$-dependent phase for generic $\Omega_B$, and equal to a random $O(1)$ phase for $\Omega_B ={\bf 1}$. 
We now collect all terms and consider the four  qualitatively different cases.  

(i) For a pure stationary  (diagonal) state, corresponding to a single non-vanishing $\rho_\alpha$ and $\Omega_B ={\bf 1}$ we find
\eqn\nps{
|{\rm noise}^{(s)}|_{\rm pure\;diag} \sim |b\,b'|\,e^{-S/2}\;.
}

(ii) For a pure generic (non-diagonal) state, corresponding to generic $\Omega_B$ we find 
\eqn\npnd{
|{\rm noise}^{(s)}|_{\rm pure\;non-diag} \sim |b\,b'|\,e^{-S}\;.
}

(iii) For a highly mixed diagonal state, corresponding to all $\rho_\alpha \sim e^{-S}$ and $\Omega_B = {\bf 1}$ the result is
\eqn\nmd{
|{\rm noise}^{(s)}|_{\rm mixed\;diag}  \sim |b\,b'| \, e^{-S}\;.
}

(iv) For a highly mixed non-diagonal state, corresponding to all $\rho_\alpha \sim e^{-S}$ and generic $\Omega_B$, we have 
\eqn\nmnd{
|{\rm noise}^{(s)}|_{\rm mixed\;non-diag} \sim |b\,b'| \,e^{-3S/2}\;.
}

We can summarize the noise phenomenology of the smooth component by saying that the mixed-aligned case has the same
noise level as the pure-misaligned case, both proportional to $e^{-S}$. Furthermore, mixed states have lower noise by a factor 
of $e^{-S/2}$ with respect to the analogous pure state, and misaligned states also have a noise level down by a factor of $e^{-S/2}$ with respect to their aligned counterparts. This means that the states with higher `smooth' noise level, of  order $e^{-S/2}$ are the exact energy eigenstates. Those  with lower noise level, of order $e^{-3S/2}$  are the highly mixed, time-dependent states.

We now turn to the estimate of the `erratic' piece induced by the second term in \corp. In this case, we have an uncorrelated choice of phases for the
operators $B(t)$ and $B'(0)$, so that their product $B(t)B'(0)$ has the general ETH form \ethband\ with non-vanishing diagonal terms. 
The  rotation of such an ETH matrix by a generic unitary $\Omega_B$ gives another  ETH matrix of the same type. So we find 
\eqn\erragen{
G_{BB'}^{(e)} (t) \sim |b\,b'|\, e^{-S/2} \sum_\alpha \rho_\alpha \left(\Omega_B^\dagger \,R_{BB'} \, \Omega_B \right)_{\alpha\alpha}
\sim |b\,b'|\, e^{-S/2} \sum_\alpha \rho_\alpha  \,\left(R_{\Omega^\dagger BB' \Omega} \right)_{\alpha\alpha}\;,
}
quite independently of the choice of $\Omega_B$. For a pure state   the correlation  is of order
$|b b'| e^{-S/2}$. For a highly mixed state the final sum over the $\alpha$ index  is proportional to $e^{-S} \Tr R \sim e^{-S/2}$, which gives an overall noise level of order $|b\,b'| \,e^{-S}$. 

Hence, the alignment (stationarity) does not affect the erratic contribution to the noise, which is in all cases larger or equal to the
smooth component. In view of our general remarks regarding the ambiguities in the separation of noise components, we interpret the erratic estimate as the true noise level of the correlation function. In this case, we find the following general rule 
\eqn\noiseoneside{
|{\rm noise}|_{\rm mixed} \sim |b\,b'| \,e^{-S}\;, \qquad 
|{\rm noise}|_{\rm pure} \sim |b\,b'|\,e^{-S/2}\;. 
}
for the overall level of Bob's noise. 
In both cases, the quasinormal relaxation time becomes of order $t_{\rm noise} \sim \Gamma^{-1} \,S$.

 \subsec{EPR Noise}
 
 \noindent
 
The general EPR correlation takes the form 
\eqn\otravez{
G_{AB}(t) = \sum_{\alpha\beta} \sqrt{\rho_\alpha \rho_\beta} \,\left(\Omega_A^\dagger \,A(t) \,\Omega_A \right)_{\alpha\beta} \left(\Omega_B^\dagger \,B(0)\,\Omega_B\right)_{\alpha\beta}\;. 
}
We shall focus on the highly entangled case, corresponding to all $\rho_\alpha \sim e^{-S}$, since the
unentangled correlation splits into the product of two one-point functions.

 We begin by estimating  the smooth component. Using \sch, \ethband\  and \corpa\ one  finds
\eqn\smoogen{
G_{AB}^{(s)} (t) \sim |a\,b| \,e^{-S}\, \left|\sum_n g_{nn} \,e^{-iE_n t} \right|^2\;.
}
where $g_{nn}$ has  the  general form 
$$
g_{nn} = \sum_\alpha (\Omega_B)_{n\alpha} \,\sqrt{\rho_\alpha}\,(\Omega_A)_{n\alpha} \sim e^{-S/2} \sum_\alpha (\Omega_B)_{n\alpha} \,(\Omega_A)_{n\alpha} \;.
$$
For a diagonal state, corresponding to $\Omega_A = \Omega_B ={\bf 1}$, we find $g_{nn} \sim e^{-S/2}$ and positive. For
a non-diagonal state, with generic choices of $\Omega_{A,B}$, the result is $g_{nn} \sim e^{-S}$ with random phase.

With these ingredients we can analyze \smoogen, whose $n$-sum is coherent at $t=0$ provided the $g_{nn}$ has a coherent phase, and always incoherent at large times, as a result of the random phases $e^{-iE_n t}$. This meas that the $t=0$ peak is there
for diagonally entangled states, and gone for unaligned states. In this last case the value of the correlation at $t=0$ is of order $|a\,b| \,e^{-S}$.  

The noise level at large times in the smooth component is then of order $|a \,b|\,e^{-S}$ for diagonal entanglement and 
$|a\,b|\,e^{-2S}$ for non-diagonal entanglement. We see that misaligning the entanglement with respect to the energy basis brings down the overall size of the (smooth) correlation by a factor of $e^{-S}$,  a result emphasized in \refs\polmarol.

The erratic component, whereby no phase correlation is assumed between $A(t)$ and $B(0)$, can be estimated as in the single-side correlations. Given the ETH form for both $A(t)$ and $B(0)$, their rotations by $\Omega_A$ and $\Omega_B$ give new ETH operators in the Schmidt basis, with matrix entries of size $e^{-S/2}$, including diagonals.  
 Therefore, the result does not depend on the alignment, and we always find  $e^{-S}$ times the trace of a generic ETH operator, amounting to a noise level of order $|a b| e^{-S}$. 
 
 As before, we regard the erratic contribution as a better indication of the true noise level, which in any case dominates over the
 noise induced by the smooth component. Hence, we find an amplitude of  EPR noise given by 
\eqn\noisepr{
|{\rm noise}|_{\rm EPR} \sim |a\,b|\,e^{-S}\;,
}
 independently of the character  of  the entanglement.

 \subsec{Beyond Narrow Bands}

 \noindent 

Combining a series of narrow bands we can obtain generalizations of the previous noise estimates. For example, we can apply
  the smooth form of the ansatz \eth\ to  the erratic contribution to a same-side, aligned correlation:
 \eqn\esti{
 {\overline{|D_{BB'}(t) |^2}} \sim  \int  dE_m \,dE_n\,\beta(E_m) \,\beta(E_m)\, e^{(S(E_m) +S(E_n))} \rho(E_m)^2\, e^{-2S({\bar E})} |b({\bar E}, \omega)|^2 |b'({\bar E}, \omega)|^2\;.
 }
 Using the same approximation assumptions of section 3.1  we obtain 
 \eqn\noisg{
 \overline{|D_{BB'}(t)|^2} \sim \int d{\bar E}\, \beta({\bar E}) \,|b({\bar E})|^2\,|b'({\bar E})|^2 \,\rho({\bar E})^2\;.
 }
 The absence of any remaining entropy factors is the most salient feature of this formula. Barring an explicit tuning of the bare operator amplitudes $b({\bar E})$ and $b'({\bar E})$, we see  that a canonical state with $\rho(E) \propto \exp(-\beta E) $
 will have noise  dominated by very low energies of order $1/\beta$. This suggests  that low energy bands have a tendency to give the largest contribution to the noise, even if they may not dominate de thermodynamic functions. 
 
A useful parametrization of the relative noise contribution of different chaotic bands is obtained by modeling a spectrum by a set of narrow bands
 and an overall quantum state:
 \eqn\overst{
 \rho = \sum_b \,p_b\,\rho_b = \sum_b \,p_b \,e^{-S_b} {\bf 1}_b\;, \qquad \sum_b p_b =1\;,
 }
 i.e. we take a maximally mixed state on each band, and assign probability weights $p_b$ to the different bands. 
 Under the general ETH assumptions for the operators, correlation functions split into a sum over the bands:
 \eqn\corrsp{
 C(t) \approx \sum_b p_b\, C_b (t)\;,
 }
 where
 \eqn\bandcc{
 C_b (t) \sim |{\rm noise}|_b \,f_b (t)\;.}
 In this expression, $f_b (t)$ is an $O(1)$ function taking into account the detailed oscillation structure of the correlation, while
 $|{\rm noise}|_b$ gives the average noise amplitude  at generic times. Using our main result $|{\rm noise}|_b \sim |bb'|\,e^{-S_b}$ we find
 \eqn\cohes{
 C(t) \sim \sum_b |bb'|_b \,p_b \,e^{-S_b} \,f_b (t)\;.}
 
 This result indicates that chaotic bands with larger value of $p_b \,e^{-S_b}$ will make the dominant contribution to the average noise amplitude. An important  particular case is the  canonical 
  weighting with inverse temperature parameter $\beta$: 
 \eqn\canw{
 p_b \Big|_{\rm canonical} = {e^{-I_b (\beta)} \over Z(\beta)}\;,}
 where $Z(\beta) \equiv \sum_b e^{-I_b (\beta)}$ and $I_b(\beta) = \beta E_b -S_b$. In this case, the overall contribution of each band to the noise amplitude is proportional to $\exp(-\beta E_b)$, confirming the dominance rule of the lowest chaotic bands.

 \subsec{Thermal Gas Noise}

 It is interesting to see what levels of noise we can expect when the ETH chaotic assumption does not hold. An idealized example of this kind is a free thermal gas, which makes a natural appearance as a model of the graviton-gas phases in AdS/CFT examples. For a thermal gas of free particles in a  $d$-dimensional box of size $L$, we consider a one-particle  operator
\eqn\buno{
B_1 = {1\over L^{d-1 \over 2}} \sum_s  \left(b_s \,a_s + b_s^* \,a_s^\dagger\right)\;,
}
with thermal self-correlation 
\eqn\qpar{
\bra B_1(t) B_1(0)\ket_{\rm gas} ={1\over L^{d-1}}  \sum_s \left[(1+f(\omega_s))|b_s|^2 \,e^{-i\omega_s t} + f(\omega_s) |b_s|^2 \,e^{i\omega_s t}\right]\;
+ {\rm interactions}}
The sum runs over single-particle levels with energy $\omega_s$ and thermal occupation probability   $f(\omega_s) = (e^{\beta \omega_s} -1)^{-1}$  (we consider here a bosonic operator for simplicity). The coefficients   $b_s$ are proportional to  the amplitude for the  injection or removal of one particle. A cutoff $\Gamma_B$ in the energies of such particles  is equivalent to a smearing  of the operator over length scales of order $1/\Gamma_B$. In keeping with the conventions for chaotic operators,  we can choose a cut-off $\Gamma_B \sim T$ so that  $B_1$ injects  or removes  one particle at the typical single-particle energy of the thermal gas.

  There are about $(L\omega)^d$ massless particle levels  up to energy $\omega$  on a  $d$-dimensional box of size $L$. A nontrivial noise regime arises if 
 these single-particle frequencies are rationally incommensurate among themselves. This can be achieved by slightly breaking the symmetries of the box, or by the interaction corrections.  One can then estimate 
\eqn\absaa{
 {\overline{|\bra B_1(t) B_1(0)\ket_{\rm gas} |^2}} \sim {1\over L^{2d-2}} \sum_{\omega_s < T} \left(1+2f(\omega_s) +2 f(\omega_s)^2 \right) |b_s|^4 \sim L^{2-2d}\, (LT)^{d-2}  \;,
 }
 where we have assumed massless quasiparticles so that  $|b_s|^2 \sim 1/\omega_s L$. The peak at $t=0$ scales instead as $\bra B_1^2 \ket \sim L^{1-d} (LT)^{d-1}$, so that the normalized noise amplitude scales as 
 \eqn\ratio{
 {|{\rm noise}|\over |{\rm peak}|}  \sim {1\over (LT)^{d/2}} \sim {1\over\sqrt{ S_{\rm gas}}}\;,
 }
 where $S_{\rm gas} \sim (LT)^d$ is the thermal entropy of the quasiparticle gas. 
 
These results suggest that reducing the amount of chaos in the dynamics, in the sense of departing from the maximal randomness of the ETH ansatz, has the consequence of bringing up the noise amplitude.

 \newsec{Listening To The Noise Of AdS}

 \noindent

 The estimates presented in the previous section only depend on quite general features of the system at hand. The general rule is that
 the larger the randomness in the state or the operators, the smaller the  amplitude of the noise. This suggests the possibility of interesting fine structure in the noise, revealing particular spectral patterns in the underlying system, such as the occurrence of energy bands with  vastly different density of states. 
 
  This is the case for a strongly-coupled CFT satisfying the basic conditions to have a smooth gravity dual description in AdS. 
  A very schematic characterization of such systems has a dense high energy band with energy $E_h$ and entropy $S_h$ both scaling with the central charge $N^2 \gg 1$, and a low energy band with characteristic energies $E_l \sim N^0 \ll E_h$ and correspondingly smaller entropies $S_l  \sim N^0 \ll S_h$. With the AdS/CFT case in mind, we can refer to the low-energy band as the graviton gas and the high-energy band as the large black hole band.  In many concrete AdS/CFT constructions such as the original SYM model,  new intermediate bands appear with parametric separation controlled by powers of the 't Hooft coupling  $\lambda$.  
  
  More precisely, for the $SU(N)$ SYM theory at large $N$ and large 't Hooft coupling satisfying $N\gg \lambda \gg 1$, the bulk model is ${\rm AdS}_5 \times {\bf S}^5$ with curvature radii $R$, string coupling $g_s \sim \lambda/N$,
  string length $\ell_s \sim  R/\lambda^{1/4}$ and ten-dimensional Planck length $\ell_p \sim R/N^{1/4}$. The graviton-gas band extends above the gap $1/R$ up to string-scale energy densities, or $E_{\rm Hag} \sim \lambda^{5/2} /R$. The Hagedorn band extends from $E_s$ up to the energy of a small ten-dimensional Schwarzschild black hole with stringy size, $E_{\rm sh} \sim \lambda^{-7/4} N^2 /R$. Finally, when the Schwarzschild black holes reach the size $R$, their energy is of order $E_{\rm bh} \sim N^2 /R$ and morph into the band of large AdS$_5$ black holes.

    The two black hole bands at the top of the spectrum are natural examples of chaotic dynamics, whereas the graviton-gas band at the bottom is the prime example of approximately free particle dynamics. The Hagedorn band is an interesting marginal case. Formally, we have a free string gas at $N=\infty$, but the single-string density of states is almost the same as the multi-string density of states (cf. for example \refs\ushag). This essential degeneracy suggests that a large mixing of multi-string states may be  possible even for tiny values of the string coupling. From this point of view, it seems reasonable to class the Hagedorn band as a chaotic one. The elucidation of this question is an interesting open problem.

 We  make the standard dynamical assumptions  regarding the operators. Namely, on chaotic bands we assume an ETH form with a width controlled by the microcanonical temperature. In the large $N$ limit relevant for the AdS/CFT correspondence, this width scales as   $O(N^0)$. Since the low and high energy bands are separated by typical energy differences of $O(N^2)$, there should be no  significant  mixing between the high and low energy bands within the ETH width.  On free quasiparticle bands we assume the operators to be well approximated by one-particle operators. In the AdS/CFT context, this means we consider `single-trace' operators whose multi-point functions factorize in the large $N$ limit into  products of two-point functions. For the particular case of a gauge theory these operators have the form
 \eqn\singletr{
 B\sim{1\over N} \,\Tr \,F^{\,n}
 \,
 }
 with $n\ll N$, and similar finite-degree polynomials in all the adjoint fundamental fields.  The normalization of \singletr\ is such that one-point functions are of $O(N^0)$. 
 
 The assumptions regarding matrix elements can be justified from the bulk picture on physical grounds. An operator of type 
  \singletr\ injects a bulk particle (say a graviton) from the boundary of AdS. In the low energy band dominated by quasi-free thermal gravitons, this operator has the form \buno, up to $1/N^2$ corrections. In the high-energy band, the graviton is injected into a preexisting black hole state. Under time evolution, this state represents a graviton falling to the stretched horizon in a time of order $T(E)^{-1} \log\,N^2$, where $T(E)$ is the microcanonical temperature associated to the average energy, $E$, of the original black hole state (cf. \refs\javi.)  Provided the energy of the injected graviton does not scale with $N^2$, we can neglect the back reaction of the graviton on the background and view the  full time evolution as a scrambling of the injected graviton state on the quantum horizon of the black hole. Hence, it is natural to assume that correlations of such a single-trace operator are well approximated by an ETH ansatz after a time of the order of the scrambling time. For the purposes of estimating noise, we can thus start with operators of the form \singletr, appropriately smeared over a `thermal' cell of the CFT, in such a way that they can be regarded as having the ETH form from the beginning.

With these assumptions in place we consider an approximation to the quantum state given by the ansatz \overst:
\eqn\aba{
\rho \approx \sum_b \, p_b \,\rho_b \approx \sum_b \, p_b \,e^{-S_b} {\bf 1}_b\;,
}
where ${\bf 1}_b$ is the indentity on the $b$'th band. 
Namely we have a set of microcanonical bands with relative weights $p_b$, satisfying $\sum_b p_b =1$. The correlation then splits into separate band contributions according to \bandcc:
\eqn\bc{
C(t) \approx \sum_b \,p_b \,|{\rm noise}|_b \,f_b (t)\;.
}
The detailed form of $f_b (t)$ depends on the chaotic or integrable character of the band. 
  For chaotic bands, $f_{\rm ch} (t)$ has oscillation components 
  with time step ranging from the inverse width of the band $\Delta^{-1}$, up to the minimum frequency $\Delta E_{\rm min}$ within the band. Most of the oscillations appear at around the Heisenberg time of the band, of order 
 $\Delta^{-1} \,e^{S} $, and make up the main contribution to $f_{\rm ch}(t)$. The  fast oscillations of time scale $\Delta^{-1}$ are carried by few matrix elements in the ETH operators, making smaller contributions  to $f_{\rm ch}(t)$ of size  $O(e^{-S})$.     
 
 For an integrable band the role of the Heisenberg time is played by the inverse of the average single-particle frequency, $T^{-1} \,S_{\rm gas}$,  and the fast component of quasi-period $T^{-1}$ makes a contribution to $f_{\rm gas}(t)$
 of height $1/\sqrt{S_{\rm gas}}$.  
 
We see that typical oscillations of  $f_{\rm gas}(t)$  are faster than typical oscillations of $f_{\rm ch}(t)$ because those are governed by  the effective Heisenberg time scale. The conclusion is that the more chaotic bands are revealed by looking at the longest-time oscillation structures, despite de fact that their contributions to the amplitude might be the smallest.

We can be more precise regarding the relative contribution of the different bands to the overall amplitude of the noise. Splitting \bc\ into chaotic and gas terms we can write
\eqn\splitt{
C(t) \approx \sum_{\rm ch} p_{\rm ch}\,e^{-S_{\rm ch}}  \,|b|^2_{\rm ch} \, f(t)_{\rm ch} + \sum_{\rm gas} \,p_{\rm gas}\, \left(S_{\rm gas}\right)^{-1/2} \,|b|^2_{\rm gas} \,f(t)_{\rm gas}\;.
}
Assuming that all $|b|_b$ are of the same order, the rule of band dominance depends on maximizing $p_{\rm ch} \,e^{-S_{\rm ch}}$ for chaotic bands and $p_{\rm gas}/\sqrt{S_{\rm gas}}$ for thermal gas bands. 

In carrying out comparisons with Euclidean saddle point methods, we are chiefly interested in the canonical ensemble, corresponding to the choice of band weights:
\eqn\cannn{
p_b \Big|_{\rm canonical} = |b|^2\,{e^{-I_b (\beta)} \over  Z(\beta)}\;, 
 }
 with $I_b (\beta) = \beta E_b - S_b$ and $Z(\beta) = \sum_b e^{-I_b(\beta)}$. Then, as explained in the previous section, the contribution of chaotic bands to the noise is dominated by the lowest one, with a noise of size $Z(\beta)^{-1} \;e^{-\beta E_{\rm ch}}$. 
 On the other hand, integrable bands such as a graviton gas contribute to the noise according to the thermodynamic hierarchy. Since
  the rule for noise dominance is the maximization of  $p_{\rm gas} /\sqrt{S_{\rm gas}}$ in this case, a canonical probability
  proportional to $p_{\rm gas} \sim \exp(-I_{\rm gas})$ always dominates the maximization rule for large systems. The reason is extensivity of the thermal gas phases, since an exponential of the volume always dominates over a power of the volume. 
  
  Finally, the competition between the dominant chaotic band and the dominant gas band is won by the thermal gas band. 
 Here we must compare  $p_{\rm ch} e^{-S_{\rm ch}}$ with
 $p_{\rm gas} / \sqrt{S_{\rm gas}}$.  Taking the ratio for  the $p_b$ values of a canonical ensemble 
 \eqn\canrat{
 {p_{\rm ch} \,e^{-S_{\rm ch}} \over (p_{\rm gas} /\sqrt{S_{\rm gas}})} = e^{-\beta (E_{\rm ch} - E_{\rm gas})} \,e^{-S_{\rm gas}} \sqrt{S_{\rm gas}} \ll 1\;,
 }
 where the last inequality follows from $E_{\rm ch} \gg E_{\rm gas}$. We see that black hole bands {\it never} dominate the noise in the canonical ensemble, no matter what the temperature might be. The noise amplitude is 
 \eqn\noisecc{
 |{\rm noise}|_{\rm canonical} \sim |b|^2\,{e^{-|\Delta I|} \over \sqrt{S_{\rm gas}}}\;,
 }
 where $\Delta I $ is the difference between the free energies of the dominating gas band and whatever band dominates the whole thermodynamics of the system. 
 
 For the basic  AdS/CFT model we have a Hawking--Page phase transition at $T=T_c \sim 1/R$. For $T<T_c$ we have $\Delta I \sim 0$ 
 and the noise amplitude is of $O(1)$ in the large $N$ limit. 
 Indeed, not even the suppression by $1/\sqrt{S_{\rm gas}}$ kicks in significantly, since the Hawking-Page temperature is right at the scale of the gap,  $T_c \sim 1/R$, and $S_{\rm gas} \sim O(1)$ just below the critical temperature. \foot{Genuine gas phases can be engineered in  confining models, where we can parametrically separate the scale of the critical temperature $T_c $ from the volume-induced scale $1/L$. In this way we can  have a large gas entropy $S_{\rm gas} \sim (LT)^{d-1}$ for $1 \ll TL < T_c L$. }

In the  high temperature regime  $T\gg T_c$ the thermodynamics is dominated by large AdS black holes, but 
the noise has amplitude \noisecc\ with 
 $|\Delta I| \sim  N^2 (RT)^{3}$ and $S_{\rm gas} \sim (RT)^9$.  The contribution of each chaotic band at $T\gg T_c$ is an additive correction to \noisecc\ with   overall amplitude $
 e^{-\beta E_{\rm ch}} /Z(\beta)$.  Hence, it is largest for the Hagedorn band (if regarded as chaotic) and smallest for  the large black hole band. In view of \noisg, the noise contribution of each chaotic band is dominated by its bottom energy scale. For the basic AdS/CFT model we then find
 $$
 |{\rm noise}|_{\,T\gg T_c} \sim |b|^2\,e^{I_{\rm bh}} \left[{1\over (RT)^{9/2}} + O\left(e^{-c_{\rm Hag} \,\lambda^{5/2}}\right) + O\left(e^{-c_{\rm sh} \,N^2 /\lambda^{7/4}}\right) + O\left(e^{-c_{\rm bh} \,N^2}\right) \right]\;,
 $$
 where $I_{\rm bh} \sim -N^2 (RT)^3$ and the positive $O(1)$ constants $c_b$ make reference  to the corresponding chaotic bands.  We see that the dominant noise is given by the $1/\sqrt{S_{\rm gas}}$ term, albeit suppressed by the inverse of the canonical partition function. Large AdS black holes dominate this partition function for $T\gg T_c$,   despite the fact that they  make the smaller contribution to the average noise amplitude (cf. figure 4).

  \bigskip
\centerline{\epsfxsize=0.6\hsize\epsfbox{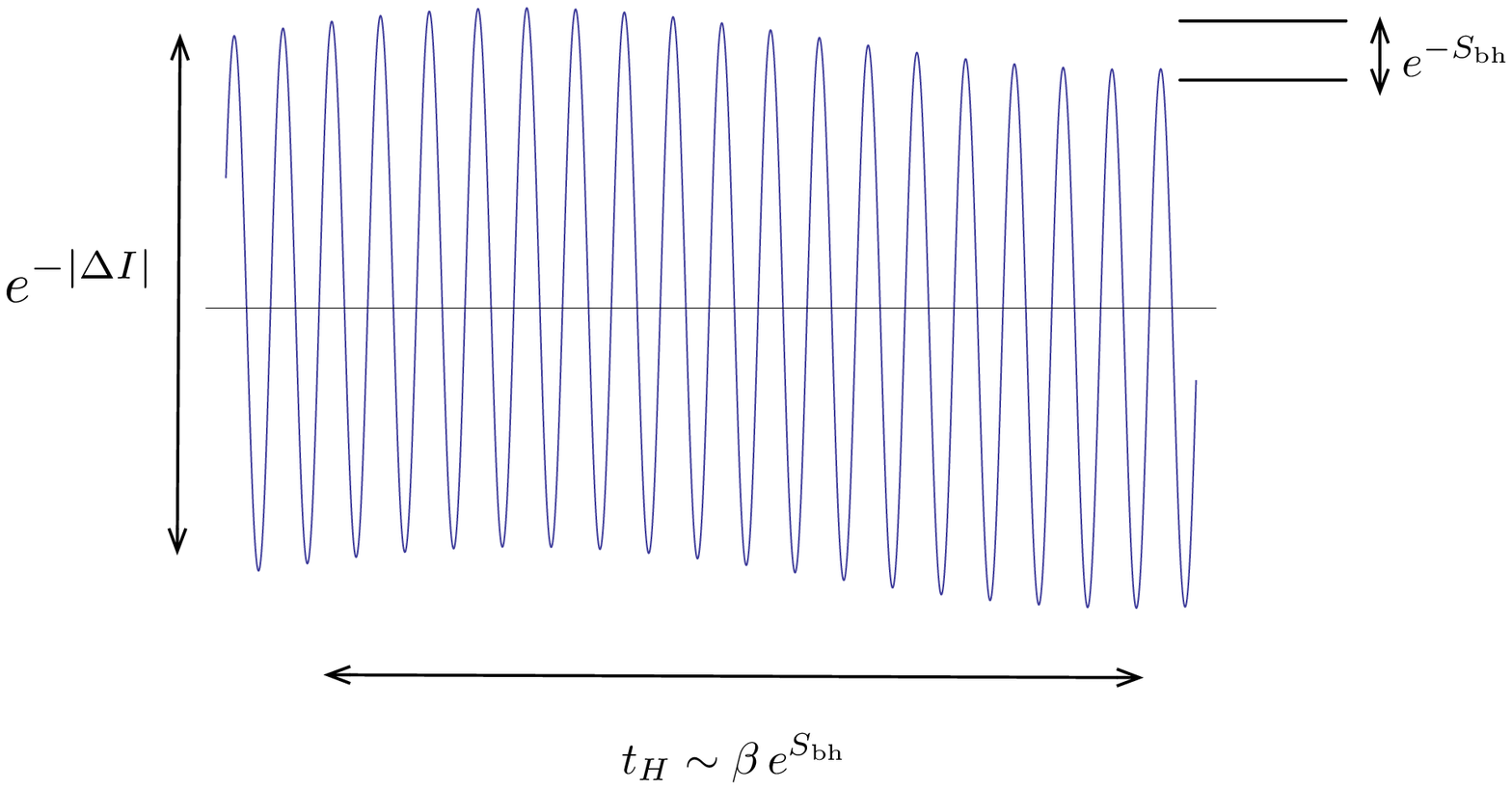}}
\noindent{\ninepoint\sl \baselineskip=2pt {\bf Figure 4:} {\ninerm
Detail of the quantum noise above the Hawking--Page transition,  on time scales of the order of the Heisenberg time of the black hole band  $t_H \sim \beta e^{S_{\rm bh}}$. The longer wavelength modulation of small amplitude corresponds to the black hole states.  The main contribution to the noise  amplitude oscillates with the faster quasi-period of the graviton gas and dominates despite the thermodynamic suppression since $e^{-|I_{\rm bh}|} \gg e^{-S_{\rm bh}}$.  
}}
\bigskip

 The noise dominance by the graviton band is not really a consequence of the large noise amplitude typical of integrable systems. Even if we allow the graviton band to interact chaotically, its noise level is suppressed by $
\exp(-S_{\rm gas})$ and the ratio analogous to  \canrat\ implies that $p_{\rm bh} \,e^{-S_{\rm bh}} \ll p_{\rm gas} \,e^{-S_{\rm gas}}$ for
the canonical state, ensuring that the noise is still dominated by the graviton gas at all temperatures. Above the Hawking--Page transition such a chaotic graviton gas would contribute an overall noise amplitude of order $e^{-|\Delta I|} e^{-S_{\rm gas}}$.

 As advanced in the previous section, we  learn that the condition for dominance of chaotic high-energy bands  is to deplete sufficiently the low-energy  bands, well below its canonical equilibrium population. We can interpret such states as `superheated', since they would settle into a lower temperature canonical state  if put in contact with a reservoir.

In a system with different bands, a hierarchy of Poincar\'e times exists, scaling with a double exponential of the corresponding entropy, and governing the recurrences within each band contribution for correlations splitting as 
in \bc.  We have seen that time averages of the noise are weighted by the effective factor $p_b \,e^{-S_b}$. Peak values at a given instant in time are however weighted by the standard state probability factor $p_b$. Hence, peak values are controlled by the
 band which dominates the thermodynamics of the system. This means that $O(1)$ recurrences of a correlation at high temperature in the AdS/CFT models will be dominated by the large black hole band and occur on time scales of order $E_{\rm bh}^{-1} \exp\left(e^{S_{\rm bh}}\right)$.

\newsec{Discussion}

\noindent

In this paper we have revisited the problem of estimating quantum noise levels in correlation functions of AdS/CFT systems, updating previous estimates made in \refs{\us, \more, \oldsus}, taking into account the natural assumption that black hole states undergo chaotic dynamics.  We use the ETH ansatz to model such chaotic dynamics and find some general results. 

The fluctuation amplitude of correlations at long times is of order $e^{-S}$ for highly mixed states over a chaotic band of $e^S$ states and of order $e^{-S/2}$ for generic pure states over the same band (see \refs\typi\ for other examples of a similar  exponential hair-splitting.) The analogous noise level for quasiparticle gas systems in found to scale like $1/\sqrt{S}$. 

These results were obtained for a narrow band whose width is of the order of the effective temperature $T$. Combining bands with different effective entropies we can analyze systems with `deconfining' phase transitions such as the standard AdS/CFT models. If $p_b$ is the thermodynamic  weight of a given band, we find that the overall noise amplitude is determined by the maximization of $p_b \,e^{-S_b}$ among chaotic bands and
$p_b /\sqrt{S_b}$ among quasiparticle bands. It follows that the noise amplitude in the canonical high-temperature regime of AdS/CFT systems is still dominated by the low-lying graviton-gas band, since the thermodynamic suppression is still larger than the black noise level: $e^{-|I_{\rm bh}|} \gg e^{-S_{\rm bh}}$. This result  agrees with the bulk instanton approximation of \refs\malda, despite the fact that this method misses completely the noise in black hole states. 
We also note, in agreement with \refs\us, that Poincar\'e recurrences are controlled by $p_b$ alone, and thus cannot be correctly characterized within the bulk instanton approximation.

We have also analyzed  EPR correlations in the purified doubled  system. Here we find   a noise level of order $e^{-S}$ for purifications with maximal entanglement, independently of the degree of alignment of the entanglement. Hence, while
any large EPR correlation  peaks  are wiped out by a misalignment of the entanglement, as emphasized in \refs\polmarol, we find that the asymptotic noise level is rather insensitive to the misalignment. 

It is unclear if this result has a bearing on the generality of the EPR=ER conjecture beyond aligned entangled states. A better proxy for such a diagnostic would be the spontaneous  emergence of $O(1)$ correlation peaks over long periods of time. For this to happen within the ETH class of operators, there must be hidden phase correlations between the Alice and Bob operator algebras.  
In the language of section 2 of this paper, a measure of the misalignment is given by the generic character of the matrices $\Omega_A$ and $\Omega_B$ which determine the Schmidt basis of Alice and Bob.    An $O(1)$  correlation peak at $t=0$ requires
ETH operator constraints of the form 
  \eqn\alicess{
 \left(\Omega_A^\dagger \,R^A\,\Omega_A\right)_{\alpha\beta}  \left(\Omega_B^\dagger \,R^{B}\,\Omega_B\right)_{\gamma\delta} \sim
 \delta_{\alpha\gamma} \,\delta_{\beta\delta} + ({\rm erratic})_{\alpha\beta\gamma\delta}\;
 }
 on a narrow band of states. Transforming this condition back to the energy basis we find 
 \eqn\eprccc{
 (R^A)_{mn} \,(R^B)_{rs} \sim (\Omega_A \,\Omega_B^T)_{mr} \,(\Omega_A^* \,\Omega_B^\dagger)_{ns} + ({\rm erratic})_{mnrs}\;,
 }
 an expression which depends non-trivially on the matrix $\Omega_A \Omega_B^T$ and looks quite intricate.  Let us consider for simplicity the case of a microcanonical Bob-side density matrix, corresponding to all $\rho_\alpha = e^{-S}$ on a band of $e^S$ states. Then, the entanglement misalignment is precisely measured by the combination  $\Omega = \Omega_A \Omega_B^T$, and the form of \eprccc\ suggests  a  generalization of  \corpa\ to 
 \eqn\corpaomega{
 (R^A)_{mn} \,(R^B)_{rs} = \Omega_{mr} \,\Omega^*_{ns} \,(\CG_{AB})_{mn}  + ({\rm erratic})_{mnrs}\;,
 }
 with smooth $(\CG_{AB})_{mn}$. This formula determines the phase correlations satisfied by those Alice operators that produce  `EPR clicks' with a given Bob operator. 
 When inserted back into the EPR correlation \eprc\ we find 
 \eqn\smotune{
 \bra A(t)\,B(0)\,\ket_\Omega \sim e^{-2S} \sum_{mn} (a\,b\,\CG_{AB})_{mn} \,e^{-i(E_m -E_n)t} + O({\rm noise}) \;,
 }
  which has the properties of a smooth $O(1)$ correlation peak with a plausible geometrical interpretation along the lines of section 3.1. The dependence on the entanglement misalignment matrix $\Omega$ has been pushed to the noise terms, which are non-perturbative from the point of view of any putative geometrical description. As we have seen in this paper, the overall amplitude of those noise fluctuations is also independent of $\Omega$.

 The upshot of this discussion is that generic-state  EPR correlation peaks with plausible geometrical interpretation require a state-dependent version of the ETH ansatz  \corpaomega. It would be interesting to elucidate whether such tuning of the operator algebra can become dynamical in some situations, as implied in \refs{\statedepv, \statedepp}. At any rate, once a peak is established at some value of the time variables, the Poincar\'e recurrences ensure an infinite landscape of such peaks. 

 {\bf Note added:} While this paper was being prepared for publication we received \refs\ellos, which contains some overlap with section 4.
 
\bigskip{\bf Acknowledgements:} 
We whish to thank  J. Mag\'an and J. Simon for many discussions over the years on the subject of this paper. We also  thank  J. Polchinski  for very valuable discussions and correspondence on the topics appearing in this work,  and for
his invaluable advice on what does not appear in the paper.

The work  of J.L.F. Barb\'on was partially supported by MEC and FEDER under a grant  FPA2012-32828, the Spanish
Consolider-Ingenio 2010 Programme CPAN (CSD2007-00042), Comunidad Aut\'onoma de Madrid under grant HEPHACOS S2009/ESP-1473 and the 
spanish MINECO {\it Centro de excelencia Severo Ochoa Program} under grant SEV-2012-0249. 

The work of E. Rabinovici  is partially supported by the
American-Israeli Bi-National Science Foundation,  the Israel Science Foundation Center
of Excellence and the I Core Program of the Planning and Budgeting Committee and The Israel 
Science Foundation ``The Quantum Universe". Eliezer Rabinovici also wishes to thank the support of the  {\it Simons Distinguished Professor Chair at KITP}, and the {\it Chaires Internationales de Recherche Blaise Pascal},  financ\'ee par l' Etat et la Region d'Ille-de-France, g\'er\'ee par la Fondation de l'Ecole Normale 
Superieure.

{\ninerm{
\listrefs
}}

\bye

\end